\documentclass[numbers]{elsarticle}
\usepackage{latexsym}
\usepackage{natbib}
\usepackage[latin2]{inputenc}
\usepackage[hidelinks]{hyperref}
\usepackage{longtable, lineno}

\makeatletter
\def\ps@pprintTitle{%
	\let\@oddhead\@empty
	\let\@evenhead\@empty
	\def\@oddfoot{\centerline{\thepage}}%
	\let\@evenfoot\@oddfoot}
\makeatother

\begin{document}

\title{The Robustness and the Doubly-Preferential Attachment Simulation of the Consensus Connectome Dynamics of the Human Brain}


\author[p]{Balázs Szalkai}
\ead{szalkai@pitgroup.org}
\author[p,u]{Vince Grolmusz\corref{cor1}}
\ead{grolmusz@pitgroup.org}
\cortext[cor1]{Corresponding author}
\address[p]{PIT Bioinformatics Group, Eötvös University, H-1117 Budapest, Hungary}
\address[u]{Uratim Ltd., H-1118 Budapest, Hungary}

\date{}


\begin{abstract}
The increasing quantity and quality of the publicly available human cerebral diffusion MRI data make possible the study of the brain as it was unimaginable before. The Consensus Connectome Dynamics (CCD) is a remarkable phenomenon that was discovered by continuously decreasing the minimum confidence-parameter at the graphical interface of the Budapest Reference Connectome Server (\url{http://connectome.pitgroup.org}). The Budapest Reference Connectome Server depicts the cerebral connections of $n=418$ subjects with a frequency-parameter $k$: For any $k=1,2,...,n$ one can view the graph of the edges that are present in at least $k$ connectomes. If parameter $k$ is decreased one-by-one from $k=n$ through $k=1$ then more and more edges appear in the graph, since the inclusion condition is relaxed. The surprising observation is that the appearance of the edges is far from random: it resembles a growing, complex structure, like a tree or a shrub (visualized on \url{https://www.youtube.com/watch?v=yxlyudPaVUE}).  Here we examine the robustness of the CCD phenomenon, and we show that it is almost independent of the particular choice of the set of underlying individual connectomes, demonstrating the Consensus Connectome Dynamics. This result shows that the CCD phenomenon is very likely a biological property of the human brain and not just a property of the data sets examined.  We also present a simulation that well-describes the growth of the CCD structure: in our random graph model a doubly-preferential attachment distribution is found to mimic the CCD: a new edge appear with a probability proportional to the sum of the degrees of the endpoints of the new edge. We hypothesized earlier that the CCD phenomenon was a good description of the axonal development of the brain on a macroscopic level: the new axonal connections prefer connecting to neurons with numerous existing connections; the here demonstrated success of the doubly-preferentially attached model is strongly in line with this assumption.
\end{abstract}

\maketitle

\section*{Introduction} 

The plenty of high-quality structural MRI data from the human brain makes possible of studying the cerebral anatomy in an unprecedented way today. Among other large projects, the NIH-funded Human Connectome Project \cite{McNab2013} records and publishes multimodal MRI data from hundreds of healthy individuals. Diffusion tensor imaging (DTI) data \cite{Hagmann2008} can be processed to discover connections, consisting of axonal fibers, between anatomically identified \cite{Fischl2012} gray matter areas. Consequently, a braingraph or connectome can be constructed that contains the connections as follows: the nodes or vertices of the graph correspond to the anatomically identified gray matter areas, and two nodes are connected if fibers of axons are discovered between them by processing the DTI data \cite{Daducci2012,Gerhard2011,Tournier2012}. 

If we have braingraphs from several hundred subjects, then, since the vertices of the different braingraphs are corresponded to the very same brain map \cite{Fischl2012},  we can describe the diversity of the edges between different subjects and in different lobes or smaller brain areas as in \cite{Kerepesi2015c}, or we can just describe the common edges through numerous subjects, as in the Budapest Reference Connectome Server \url{http://connectome.pitgroup.org} \cite{Szalkai2016,Szalkai2015a}. The data source of these studies was the Human Connectome Project \cite{McNab2013}.

Distinguishing the frequently and rarely appearing connections within the human brain may help the neuroscientist in identifying the normally appearing, usual, standard and non-standard connections. These non-standard connections can cause or can be caused by some disease, or just can be the result of the personal variability with or without psychological consequences. Therefore, the mapping of the frequent and the infrequent connections by the Budapest Reference Connectome Server can have straightforward clinical significance.

Very surprisingly, we have discovered a phenomenon, called Consensus Connectome Dynamics (CCD), on the Budapest Reference Connectome Server, which may open up new horizons in the study of the development of the human brain \cite{Kerepesi2015b}. This discovery was very surprising since the server was not built to study brain development: the imaging data is originated from adults between 22 and 35 years of age, so the age-span is --- seemingly --- inadequate for studying the early development of brain connections, occurring in several months just before and after the birth \cite{Lewis2013}. To clarify this discovery, we need to cover some details of the Budapest Reference Connectome Server \url{http://connectome.pitgroup.org}.

The server is capable of computing and visualizing the consensus connectomes with setting several parameters. The braingraphs of $n=418$ subjects are processed on the server. Let $k$ be an integer such that $1\leq k \leq n$. Let us call an edge $e$ $k-frequent$ if the edge $e$ is present in at least $k$ braingraphs out of the maximum $n$ graphs. Let us call a connectome $k$-consensus connectome if it contains all the $k$-frequent edges. The $k$-consensus connectomes, consequently, contains all edges that are present in at least $k$ connectomes. For $k=1$, the 1-consensus connectome contains all the edges that are present in at least one subject's braingraph out of the $n$ graphs. The $n$-consensus connectome contains the edges that are present in all subject's connectomes. Clearly, the $n$-consensus connectome contains much fewer edges than the 1-consensus connectome. Let $1\leq i \leq  j \leq n$ then it is also obvious that the edges of the $j$-consensus connectome are also present in the $i$-consensus connectome. This means that the $i$-consensus connectome contains more and more edges by the decrease of $i$.

The fascinating observation \cite{Kerepesi2015b} is that the new edges that appear in the -- larger -- $i$-consensus connectome, relative to the -- smaller -- $i+1$-consensus connectome, are not placed randomly, they seem to connect to the edges of the $i+1$-consensus connectome. Consequently, if we consider the $k$-consensus connectomes, for decreasing $k$ values from $k=n, n-1, ...,3,2,1$ then we get more and more edges, and the edges form a growing, complex, tree-like structure. 

The observation is visualized on a video at \url{https://youtu.be/yxlyudPaVUE} for the whole brain and at \url{https://youtu.be/wBciB2eW6_8} restricted for the frontal lobe only  \cite{Kerepesi2015b,Kerepesi2016}. The observation is statically visualized on a very large component-tree at  \url{http://pitgroup.org/static/graphmlviewer/index.html?src=connectome_dynamics_component_tree.graphml}, which is described in detail in \cite{Kerepesi2015b}. The observation is analyzed quantitatively on Figure 2 in \cite{Kerepesi2015b} for the whole brain and on Figure 1 of \cite{Kerepesi2016} for the frontal lobe only.

The interested reader can also experience the Consensus Connectome Dynamics phenomenon on the Budapest Reference Connectome Server \url{http://connectome.pitgroup.org} by (i) choosing the ``Show options'' button and (ii) moving the ``Minimum edge confidence'' slider to the rightmost position, and (iii) slowly moving the ``Minimum edge confidence'' slider from right to left.

In \cite{Kerepesi2015b,Kerepesi2016} we hypothesized that the Consensus Connectome Dynamics (CCD) phenomenon describes the order of the development of the connections of the brain: the deviation of the oldest, first developed connections are the smallest, and, gradually, the newer and newer developed connections cumulate the deviations of the connections that they connect to, and, because of this, their deviation will be higher and higher; that is, they will appear only in $k$-consensus connectomes for larger values of $k$.

\section*{Discussion and Results}

In the present contribution we examine two relevant questions concerning the CCD phenomenon:

\begin{itemize}
	
	\item[a:] {\bf Robustness:} The CCD phenomenon can be characterized by the order of appearance of the edges in the growing graphs of $k$-consensus connectomes, for the decreasing $k$ parameters. For showing that this order has any biological meaning and it is not just the product of the specific choice of the dataset processed, we need to demonstrate the independence of the appearance of the edges from the particular choice of the underlying dataset.	
	
	\item[b:] {\bf Random graph model for CCD:} The main reason for preparing a random network model for a known, interesting graph is uncovering the possible mechanism involved in the development (or evolution) of the graph. As the most famous example, the Barabási-Albert model for the description of the degree distribution of the webgraph \cite{Barabasi1999,Bollobas2001} uses the ``preferential attachment'' principle in the description of the development of the webgraph. In the  Barabási-Albert model, roughly, in every step one new vertex appears, and it connects to the older ones with probabilities proportional to the degree of the older vertices. It was shown first by computer simulation \cite{Barabasi1999} and, later by an exact mathematical proof \cite{Bollobas2001} that this process led to the power law degree distribution with exponent -3. Most importantly, since the random simulation process well-described the degree-distribution of the webgraph, the model uncovers also the mechanism that guides the users of the web in hyperlinking the new vertices (web pages) in the webgraph.
	
\end{itemize}

In what follows we show that the CCD phenomenon is robust in the sense described above, and we also define a probabilistic graph model with a ``doubly preferential attachment'' that well-describes the CCD phenomenon.

\subsection*{Robustness} 

Here we examine the independence of the CCD phenomenon of the particular choice of the underlying datasets of braingraphs. For this goal, we partitioned the 418 braingraphs into 4 disjoint sets of almost equal size with a $\pm 1$ margin, and we computed the order of appearance of the edges in the CCD, according to each dataset. Next, we have compared the order of appearance of the edge pairs as follows:

We have chosen those edges that are present in all the four CCD graphs; there are 31,873 such edges. Then we - randomly - chose two of the braingraph datasets out of the four, say $X$ and $Y$, and also randomly two edges, say $e$ and $f$ out of the 31,873. Next, we compute the experimental probability that in the $X$ dataset-based CCD edge $e$ appears strictly before $f$ and in the $Y$ based experiment $f$ appears strictly before $e$. 

If in the CCD the edges just appeared randomly this probability would be equal to 1/2. The smaller is the probability, the more robust is the CCD phenomenon. We have got this probability to be 0.104.

Similarly, we have also computed the order of the connections of the vertices to the consensus connectomes, and for a randomly chosen $u$, $v$ vertex-pair we computed the experimental probability that in the $X$ dataset-based CCD vertex $u$ appears strictly before $v$ and in the $Y$ based experiment $v$ appears strictly before $u$. We have got that this probability is 0.053. 

Therefore, we can conclude that for the edges and the vertices, the order of their appearance in the CCD phenomenon is almost independent of the underlying dataset, so, by our opinion, this order describes a property of the brain, and not of the datasets.

\subsection*{Random graph model for the CCD} 

There are three significant differences --- relative to the webgraph-simulation --- that need to be addressed in developing the random graph model for the CCD phenomenon:

\begin{itemize}
	
\item{(i)} In CCD, we have the data of the buildup of the graph; in other words, in CCD we have a dynamic process of the appearances of the new edges, while in the case of the webgraph only a static image: the degree distribution of the vertices in the graph;
	
\item{(ii)} In CCD we observe new edges between those ``old'' vertices that were connected to some edges in the previous steps (i.e., with larger $k$s in $k$-consensus connectomes). In the Barabási-Albert model, new edges are always connected to the new vertex, and they never appear between two old nodes.

\item{(iii)} We do not intend to model an unboundedly growing graph as in \cite{Barabasi1999,Bollobas2001}; our goal is to model the 1015-vertex CCD phenomenon.
	
\end{itemize} 

Here we suggest a doubly-preferential attachment probability distribution for the new edges: the probability of the appearance of a new $uv$ edge between vertices $u$ and $v$ is proportional to the sum $\deg u+\deg v$, i.e., the sum of their degrees. We call this rule ``doubly-preferential attachment'', because in the Barabási-Albert model \cite{Barabasi1999} the new vertex $u$ was connected to old vertex $v$ with a probability, proportional to $\deg v$ (the ``preferential attachment model''). The mathematical details and the parameter choices are detailed in the ``Methods'' section.

Figure 1 compares the increase of the edge numbers in the real CCD phenomenon and in the doubly-preferential attachment model we suggest. Step $i$ on the horizontal axis correspond to the $(n+1-i)$-consensus connectome. 

\begin{figure} [h!]
	\centering
	\includegraphics[width=5in]{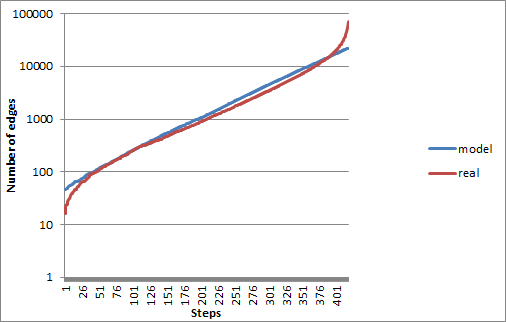}
	\caption{The comparison of the edge numbers in the CCD phenomenon with the edge numbers in the doubly-preferential attachment random model we suggest. The horizontal axis contains the numbering of the steps: step $i$ corresponds to the $(n+1-i)$-consensus connectome.}
\end{figure}

Figure 2 compares the sum of the isolated edges in the CCD and the random, doubly-preferentially attached model. An edge is called ``isolated'' in the $k$-consensus connectome, if it does not connect to any other edges, and it was not present in the $k+1$-consensus connectome. One quantitative characterization of the CCD phenomenon is the very small number of isolated edges (c.f. Figure 2 in \cite{Kerepesi2015b} and Figure 1 of \cite{Kerepesi2016}). Therefore the sum of the isolated edges is a good measure of the good characterization of the CCD phenomenon.

\begin{figure} [h!]
	\centering
	\includegraphics[width=5in]{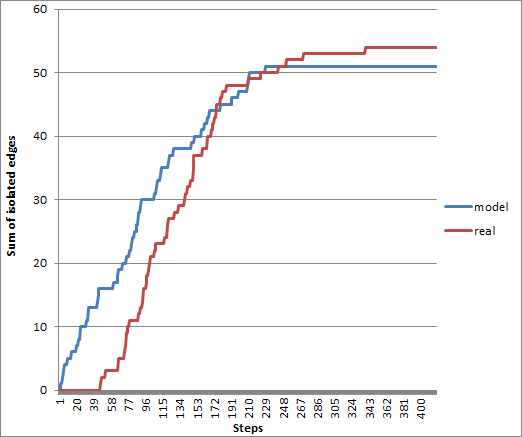}
	\caption{The comparison of the sum of the isolated edges in the CCD phenomenon with the corresponding value in the doubly-preferential attachment random model we suggest. The horizontal axis contains the numbering of the steps: step $i$ correspond to the $n-i$-consensus connectome.}
\end{figure}

\section*{Methods}

We used a Barabási-Albert-like model for approximating the connectome distribution. First, we observed that the number of edges is approximately an exponential function of $k$, with sharp increases at the beginning and at the end. An exponential regression yielded the equation $46.37 e^{0.014k}$ ($R^2 = 0.99$). Let $A := 46.37$ and $B := 0.014$. From this equation we have derived the following simple model: we start from a $\lfloor A \rfloor$-edge graph (i.e. a 46-edge graph), generated randomly over $D$ selected nodes (where $D \leq N$ is a parameter of the model) then, in each step, we add each $uv$ edge with the probability

$$
p_{uv} := \frac{B}{2(N-1)} \left( \deg u + \deg v \right),
$$

where $\deg u$ denotes the degree of node $u$ in the previous step.

This indeed yields an exponentially growing number of edges. Observe that the expected number of new edges in the next step is (if we do not account for multiple edges)

$$
\sum_{u \in V} \sum_{v \in V \atop v > u} \frac{B}{2(N-1)} \left( \deg u + \deg v \right) = \frac{B}{2(N-1)} 2 \frac{N-1}{2} \sum_{u \in V} \deg u = \frac{B}{2} 2|E| = B|E|,
$$

where $|E|$ is the number of edges in the previous step. Thus our model indeed generates an exponential expected number of edges, namely approximately $\lfloor A \rfloor e^{Bk} = 46 e^{0.014k}$ edges in the $k$th step, which is consistent with the exponential regression.

Unfortunately, this model does not allow adding new edges between zero-degree (isolated) nodes, because $p_{uv}$ becomes $0$ for those nodes. To circumvent this problem, we have modified the equation so that we allow a certain probability for the inclusion of these ``isolated'' edges. We added a constant to $p_{uv}$, that is, in our new model, the probability of inclusion for the edge $uv$ has now become

$$
p_{uv} := \frac{B}{2(N-1)} \left( \deg u + \deg v \right) + C,
$$

where $C$ is the inclusion probability for isolated edges.

This causes the number of edges to be not $\lfloor A \rfloor e^{Bk}$, but approximately $(\lfloor A \rfloor + {N\choose 2}C/B ) e^{Bk}$. This is because the expected number of new edges of $C$ is ${N\choose 2}C$ in each step, and the number of edges is multiplied by about $1+B$ in each step. Therefore, by using the $1+z+z^2+... = \frac{1}{1-z}$ formula for $z := \frac{1}{1+B}$, we can derive the number of edges relative to $e^{Bk}$. Based on this formula, we need to decrease the initial number of edges from $\lfloor A \rfloor$ to $\lfloor A - {N\choose 2}C/B + 0.5 \rfloor$ (0.5 is added for rounding to the nearest integer).

Since it is unlikely that two isolated edges appear from the same node at once, the value of $C$ influences the total number of new isolated edges in an almost linear fashion. So we can start from a value of $C$, count the total number of new isolated edges, then compare it with the desired total number of isolated edges, and divide $C$ with this ratio. This way, we determined the optimal value for C as $7.6 \times 10^{-7}$. We found that, in reality, the cumulative number of new isolated edges is 0 up until step 47, then increases in an approximately linear fashion up until step 186, and after that it levels off at 50--55 edges in total. In comparison, the average curve of the cumulative number of new isolated edges in our simulation increased linearly until about step 130, had a concave section until about step 230, where it leveled off at 56 edges.

We can thus conclude that our model for CCD not only approximates the number of edges in each step well, but also the cumulative number of isolated edges. Furthermore, to avoid over-fitting, the model is simple and only has 4 parameters.

\section*{Conclusions}

We have shown that the CCD phenomenon is robust, in other words, most probably it describes a biological  phenomenon, and it is not just the property of the particularly chosen datasets.

We have also shown that the doubly-preferentially attached model well-describes the CCD phenomenon. This fact also strengthens our hypothesis described in \cite{Kerepesi2015b,Kerepesi2016} that the CCD phenomenon copies the axonal development of the brain on a macroscopic level: there we hypothesized that the new axonal connections prefer connecting to neurons with numerous existing connections; the success of the doubly-preferentially attached model is in line with this assumption, since here new edges appear more probably between -- already -- high-degree nodes.

\section*{Data availability:} The Human Connectome Project's MRI data is accessible at: 
\url{http://www.humanconnectome.org/documentation/S500} \cite{McNab2013}. 

\noindent The graphs (both undirected and directed) that were prepared by us from the HCP data can be downloaded at the site \url{http://braingraph.org/download-pit-group-connectomes/}.

\section*{Acknowledgments}
Data were provided in part by the Human Connectome Project, WU-Minn Consortium (Principal Investigators: David Van Essen and Kamil Ugurbil; 1U54MH091657) funded by the 16 NIH Institutes and Centers that support the NIH Blueprint for Neuroscience Research; and by the McDonnell Center for Systems Neuroscience at Washington University. BS was supported through the new national excellence program of the Ministry of Human Capacities of Hungary.



\end{document}